\documentstyle[epsf,graphics,a4,12pt]{article}
\begin{document}
\newcommand{\beq}{\begin{equation}}
\newcommand{\eeq}{\end{equation}}
\newcommand{\beqn}{\begin{eqnarray}}
\newcommand{\eeqn}{\end{eqnarray}}
\newcommand{\dpf}{\displaystyle\frac}
\newcommand{\no}{\nonumber}
\newcommand{\ep}{\epsilon}
\begin{center}
{\large Object picture of quasinormal ringing on the background of small Schwarzschild
Anti-de Sitter black holes}
\end{center}
\vspace{1ex}
\centerline{\large  Jiong-Ming Zhu$^{b,}$\footnote[1]{e-mail:wlxzhu@shtu.edu.cn},
\ Bin Wang $^{a,b,}$\footnote[2]{e-mail:binwang@fma.if.usp.br}
and Elcio Abdalla$^{a,}$\footnote[3]{e-mail:eabdalla@fma.if.usp.br}
}
\begin{center}
{$^{a}$ Instituto De Fisica, Universidade De Sao Paulo,
C.P.66.318, CEP
05315-970, Sao Paulo, Brazil \\
$^{b}$ Department of Physics, Shanghai Teachers' University,
P. R. China}
\end{center}
\vspace{6ex}
\begin{abstract}
We investigated the evolution of a scalar field propagating in small Schwarzschild
Anti-de Sitter black holes. The imaginary part of the quasinormal frequency decreases
with the horizon size and the real part of the quasinormal frequency keeps nearly as a
constant. The object-picture clarified the question on the quasinormal modes for
small
Anti-de Sitter black holes. Dependence of quasinormal modes on spacetime dimensions and the
multipole index for small Anti-de Sitter black holes have also been illustrated.
\end{abstract}
\vspace{6ex} \hspace*{0mm} PACS number(s): 04.30.Nk, 04.70.Bw
\vfill
\newpage

The dynamics of radiative fields in black hole spacetimes has been investigated for
almost thirty years. It is well known that after the initial pulse, there is a
quasinormal ringing outside the black hole. The frequencies and damping times of the
``ring" are entirely fixed by the structure of the background spacetime and is believed
to be a unique fingerprint to directly identify the black hole existence. Detection of
these quasinormal modes is expected to be realized through gravitational wave observation
in the coming decade [1]. In order to extract as much information as possible from
gravitational wave signal, it is necessary to understand exactly the behavior of
quasinormal modes in different black hole spacetimes. For black holes in asymptotically
flat spacetimes, especially spherical cases, the quasinormal modes have been studied
extensively [1-7]. The study for the nonspherical black holes is developing [8-10].
Considering the case when the black hole is immersed in an expanding universe, the
quasinormal modes of black holes in de Sitter spaces have also been investigated recently
[11].

Motivated by the recent discovery of the AdS/CFT correspondence, the investigation of the
quasinormal modes of AdS black holes becomes more appealing nowdays. The quasinormal
frequencies of AdS black holes have direct interpretation in terms of the dual conformal
field theory (CFT). The first study of the quasinormal ringing for a conformally coupled
scalar in AdS space was performed by Chan and Mann [12]. Recently, 
Horowitz and Hubeny [13] considered the scalar quasinormal modes on the background of
Schwarzschild AdS black holes
in four, five and seven dimensions. They claimed that for large AdS black holes both the
real and imaginary parts of the quasinormal frequencies scale linearly with the black
hole temperature. This result was supported in a recent study
for quasinormal modes of three-dimensional BTZ black holes [14,15] and five dimensional
Schwarzschild AdS black holes [14]. Considering that the Reissner-Nordstrom (RN) AdS
solution provides a better framework than the Schwarzschild AdS geometry and may
contribute significantly to our understanding of space and time, we generalized the study
of quasinormal modes in RN AdS black holes [16,17]. We found that the charge in RN AdS
black holes showed a richer physics concerning quasinormal modes and further information
on AdS/CFT correspondence. For small AdS black holes the study is not thorough. It was
argued that quasinormal frequencies do not continue to scale with temperature in [13].
However this argument was challenged by the superpotential approach [14] where the
authors
claimed that the mode is still proportional to the surface gravity for very small AdS
holes. The available results on quasinormal modes for small AdS black holes mainly relied
on indirect arguments and focused much on quasinormal frequencies. The object-picture of
the evolution of test-field around the small AdS black hole background is lacking.

In order to clarify the clash on quasinormal mode behavior for small AdS holes, 
we will analyse in detail here the wave propagation of massless scalar field in
small Schwarzschild AdS spacetimes. We will show that the direct picture of the evolution
supports the argument of quasinormal frequencies for small AdS black holes in [13].
Whereas the temperature begins to increase as one decreases black hole horizon $r_+$
below the AdS radius $R$, we observed that it takes longer time for the perturbation to
settle down, which means that the imaginary part of the frequency continues to decrease
with $r_+$. Moreover, we will give the object-picture of the relation between quasinormal
ringing to spacetime dimensions for small AdS holes. The behavior of modes for small AdS
black holes of multipole order $l$ will also be displayed.

The $d$ dimensional Schwarzschild AdS black hole metric is given by
\beq  
ds^2=-f(r)dt^2+f^{-1}(r)dr^2+r^2d\Omega^2_{d-2}
\eeq
where
\beq       
f(r)=\dpf{r^2}{R^2}+1-(\dpf{r_0}{r})^{d-3}.
\eeq
$R$ is the AdS radius and $r_0$ is related to the black hole mass via
$M=\dpf{(d-2)A_{d-2}r^{d-3}_0}{16\pi G_d}$, where $A_{d-2}=2\pi ^{(d-1)/2}/\Gamma
(\dpf{d-1}{2})$ is the area of a unit $(d-2)$ sphere. The black hole horizon is at
$r=r_+$, the largest root of $f=0$. In this paper, we will concentrate our attention on
black holes with $r_+<R$.

Let us consider a massless scalar field $\Phi$ in the Schwarzschild AdS spacetime,
obeying the wave equation
\beq 
\Box \Phi =0
\eeq
where $\Box=g^{\alpha\beta}\nabla_{\alpha}\nabla_{\beta}$ is the d'Alembertian operator.
If we decompose the scalar field according to
\beq       
\Phi (t,r,angles)=r^{\dpf{2-d}{2}}\psi(t,r) Y(angles)
\eeq
then each wave function $\psi$ satisfies the equation
\beq       
-\dpf{\partial^2 \psi }{\partial t^2}+f\dpf{\partial}{\partial r}(f\dpf{\partial
\psi}{\partial r})=V\psi
\eeq
where
\beq   
V=f[\dpf{l(l+d-3)}{r^2}-\dpf{(2-d)(d-4)}{4r^2}f-\dpf{2-d}{2r}\dpf{\partial f}{\partial
r}].
\eeq
Using the tortoise coordinate $r^*=\int \dpf{dr}{f}$, eq(5) can be written as
\beq          
-\dpf{\partial^2 \psi }{\partial t^2}+\dpf{\partial^2\psi }{\partial r^{* 2}}=V\psi.
\eeq

The potential $V$ is positive and vanishes at the horizon, which corresponds to
$r^*=-\infty$. It diverges at $r=\infty$ corresponding to a finite value of $r^*$, which
requires that $\Psi$ vanishes at infinity. This is the boundary condition to be satisfied
by the wave equation for the scalar field in AdS space. The behavior of the potential
differs quite a lot from that of the asymptotically flat space and de Sitter space. As
argued in [18], it is this distinguishing feature that contributes to the special wave
propagation in AdS spacetimes.

Introducing the null coordinates $u=t-r^*$ and $v=t+r^*$, (7) can be recast as
\beq  
-4\dpf{\partial^2}{\partial u\partial v}\psi=V(r)\psi
\eeq
where $r$ is determined by inverting the relation $r^*(r)=(v-u)/2$.

The future black hole horizon is located at $u=\infty$. Since the quasinormal modes of
AdS space are defined to be modes with only ingoing waves near the horizon, we will pay
more attention on the wave dynamics near the event horizon.

The two-dimensional wave equation (7) can be integrated numerically,
using for example the finite difference method suggested in
[2]. Using Taylor's theorem, it is discretized as
\beq 
\psi_N=\psi_E+\psi_W-\psi_S-\delta u\delta v
V(\dpf{v_N+v_W-u_N-u_E}{4})\dpf{\psi_W+\psi_E}{8}+O(\epsilon^4)
\eeq
where the points $N, S, E$ and $W$ form a null rectangle with relative positions as: $N:
(u+\delta u, v+\delta v), W: (u+\delta u, v), E: (u, v+\delta v)$ and $S: (u, v)$.
$\epsilon$ is an overall grid scalar factor, so that $\delta u \sim
\epsilon \sim \delta v$.
Considering that the behavior of the wave function is not sensitive to the
choice of initial data, we set $\psi(u, v=v_0)=0$ and use a Gaussian
pulse as an initial perturbation, centered on $v_c$ and with
width $\sigma$ on $u=u_0$ as
\beq       
\psi(u=u_0, v)=\exp[-\dpf{(v-v_c)^2}{2\sigma^2}].
\eeq

The inversion of the relation $r^*(r)$ needed in the evaluation of the
potential $V(r)$ is the most tedious part in the computation. First 
of all, we choose a suitable value of $r$ as our starting point 
$r_0$, which should be quite close to $r_+$. It corresponds to 
the point ( $u_0, v_0$), the lowest point on the numerical grid. In the 
outer loop of our computation, $u$ increases by $\epsilon$ for every circle, 
while $v$ is kept as a constant. So we 
have $\delta r^*=-\delta u/2 \sim-\epsilon /2$ . In the inner loop, 
as $u$ is kept unchanged, we should 
take $\delta r^*=\delta v/2 \sim\epsilon /2$. Then we can use the 
relations: $\delta r=f(r)\delta r^*$ and $r=r+\delta r$ to get all the values 
of $f(r)$ and $V(r)$ at every point on the numerical grid step by step.

After the integration is completed, the value $\psi(u_{max}, v)$ are
extracted, where $u_{max}$ is the maximum value of $u$ on the
numerical grid. Taking sufficiently large $u_{max}$, $\psi(u_{max},
v)$ represents a good approximation for the wave function at the event
horizon, which carries information about the quasinormal modes for AdS 
space of our interest. We fix $R=1$ in the following.

We now report results of our numerical simulations of evolving massless scalar field on a
small Schwarzschild AdS black hole background.

For the four dimensional small black hole $(r_+<R)$, quasinormal ringings are
displayed
in Fig(1) for selected values of $r_+$ and multipole index $l=0$.

\begin{center}
\setlength{\unitlength}{1.0mm}
\begin{picture}(110,80)
\thicklines
\put(0,-2){\resizebox{110\unitlength}{80\unitlength}
{\includegraphics{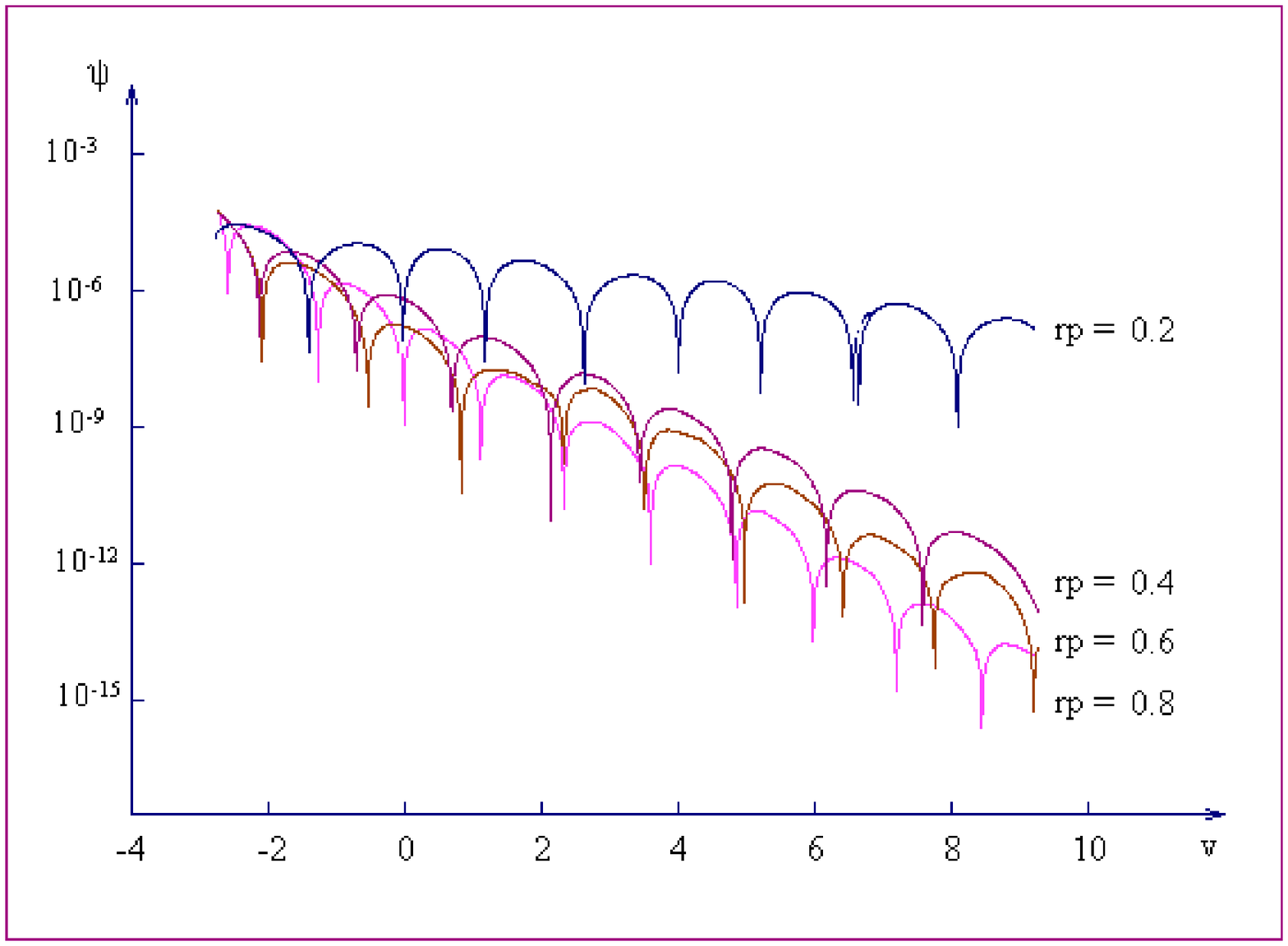}}}
\label{fig:1}
\end{picture}
\parbox[t]{\textwidth}{Figure 1: {\it The wave functions for small AdS black holes in
four-dimensions and with $l=0$}}
\end{center}

It is clear to see that the quasinormal ringing has the behavior of oscillatory
exponential falloff, which is similar to that of large AdS holes shown in [16]. This
behavior differs from that of asymptotically flat spacetime [2-7] and also de Sitter
spacetime [11,12]. According to Ching et al [18], this difference can be attributed to
the completely different boundary condition of AdS spacetime from that of asymptotically
flat and de Sitter spacetimes. The oscillation appearing in the exponential tail is due
to
the waves bouncing off the divergent potential barrier at large $r$ in AdS space.

The damping time scale and oscillation time scale of the quasinormal ringing relates to
imaginary ($\omega_I)$ and real ($\omega_R$) parts of the quasinormal frequencies. It is
interesting to note that for small AdS black holes, the evolution of the test field
experiences an increase of the damping time with the decrease of $r_+$. This corresponds
to say that imaginary part of the quasinormal frequency $\omega_I$ decreases with $r_+$
for small AdS holes. Since the decay of the test field outside the black hole is due to
the black hole absorption, it is natural to understand that when the black hole becomes
arbitrarily small, the field will no longer decay. Thus this dependence of $\omega_I$ on
black hole size for small black hole is reasonable.

When the AdS black hole becomes small enough ($r_+$ much smaller than $R=1$), from Fig(1)
we learnt that the oscillation time scales do not differ much for different small holes.
Thus the real part of the quasinormal frequencies $\omega_R$ is nearly a constant for
small AdS holes.

For five and seven dimensional Schwarzschild AdS black holes, the results are similar.
The object-picture of quasinormal ringing displayed here clarified the clash on
quasinormal modes for small AdS black holes raised in [13] and [14]. Our results support
the argument given by Horowitz and Hubeny [13]. The quasinormal modes for small AdS black
holes will not depend on the black hole temperature as was the case for large AdS holes.

In Fig(2) we show the relation between quasinormal frequencies and spacetime dimension 
for small AdS Schwarzschild black hole with $l=0$.

\begin{center}
\setlength{\unitlength}{1.0mm}
\begin{picture}(110,80)
\thicklines
\put(0,-2){\resizebox{110\unitlength}{80\unitlength}
{\includegraphics{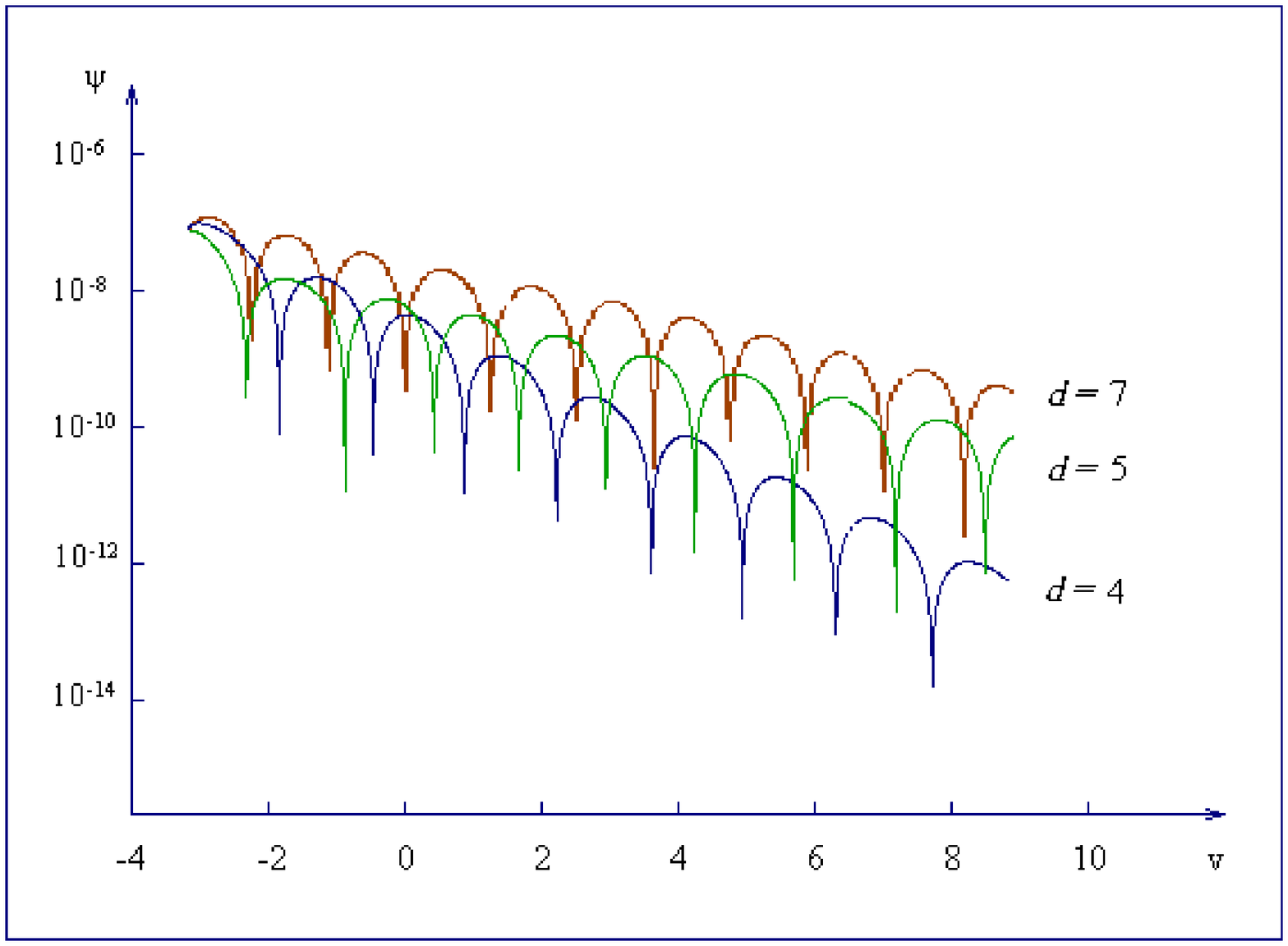}}}
\label{fig:2}
\end{picture}
\parbox[t]{\textwidth}{Figure 2: {\it The wave functions for different dimensional
small AdS black holes with $r_+=0.4, l=0$.}}
\end{center}

For big Schwarzschild AdS black holes, Horowitz et al found that $\omega_I$ is almost
independent of dimension, while in contrast $\omega_R$ does depend on the dimension [13].
In Fig(2) we observed that for small AdS black hole the $\omega_R$ keeps the same
dependence on dimension as that for big AdS hole, it increases with the increase
of spacetime dimensions. However the characteristic of $\omega_I$ for small AdS holes
differs a lot from that of big holes. We learnt that the damping time scales relate to
spacetime dimensions. The higher the dimensions are, the longer the test field
takes to settle down. This corresponds to the fact that $\omega_I$ becomes small for
higher
dimensions for small AdS holes. This property is very different from that of big AdS
black holes.

We have so far discussed only of the lowest multipole index $l=0$. Now we
show how wave dynamics behaves for the small Schwarzschild AdS background for
different multipole
index $l$.

\begin{center}
\setlength{\unitlength}{1.0mm}
\begin{picture}(110,80)
\thicklines
\put(0,-2){\resizebox{110\unitlength}{80\unitlength}
{\includegraphics{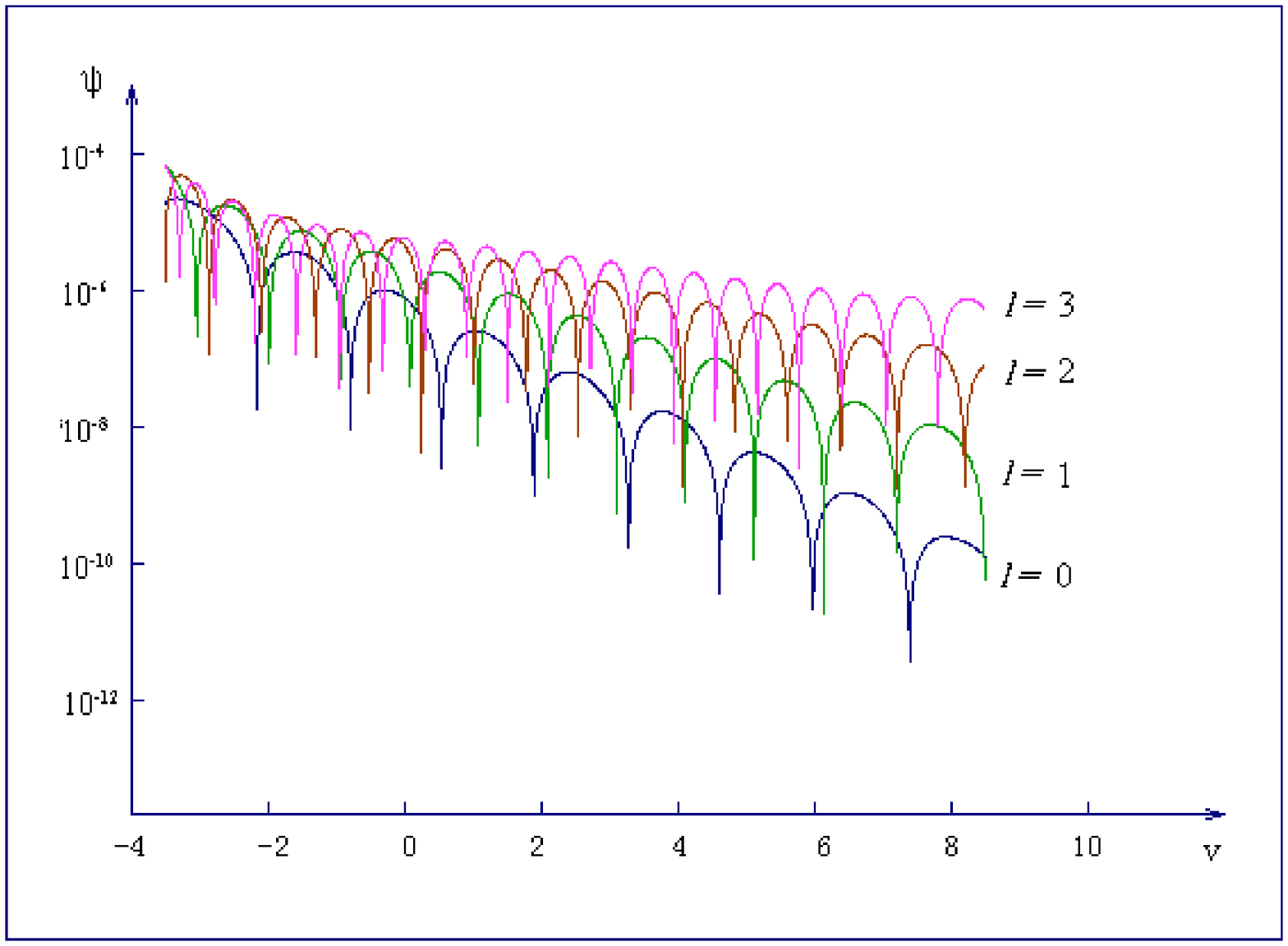}}}
\label{fig:3}
\end{picture}
\parbox[t]{\textwidth}{Figure 3: {\it Wave functions for small AdS black holes for
$r_+=0.4$, with $l=0,1,2,3$}}
\end{center}

Fig(3) exhibits a picture consistent with that given in [13,16,17] for big AdS
black holes. Increasing $l$, the evolution of the test field on the background small AdS
holes experiences an increase of the damping timescale ($\omega_I$ decreases) and a
decrease of the oscillation time scale ($\omega_R$ increases). The figure gave us
an object-lesson: the evolution of the test field with the increase of multipole
index for small AdS holes. Compared to the results given in [13], we found that
dependence of quasinormal frequencies on multipole index is universal for big,
intermediate and small AdS black holes.

In summary, we have studied the wave evolution around small Schwarzschild AdS black holes
and
revealed consistent and different results compared to that for large AdS black 
hole background. The
quasinormal ringing associated with a massless scalar field propagation on a fixed
background of small AdS black holes experiences an oscillatory exponential decay at the
black hole horizon. The smaller the AdS black hole is, the larger is the  time needed
for the
test field to settle down. However the oscillation time scales remain almost the same for
enough small AdS holes. This result supports [13]: for small AdS holes $\omega_I$
decreases with $r_+$ while $\omega_R$ keeps nearly as a constant, which is very different
from the quasinormal modes for big AdS holes where modes are proportional to black hole
temperature. The object picture we presented here clarified the clash on quasinormal
modes for small AdS black hole in [13,14] and supports the indirect argument given
in[13].

We have also learnt an object-lesson on the dependence of quasinormal ringing on
spacetime dimensions for small AdS black holes. Despite the characteristic of
$\omega_R$ similar to that of big AdS holes, it is interesting to find that $\omega_I$
for small
AdS black holes is no longer almost independent of dimension as that for big AdS hole
cases. It has been illustrated that $\omega_I$ decreases with the increase of spacetime
dimensions for small AdS holes. The consistent picture of the quasinormal modes depending
on the multipole index $l$ between small and big AdS holes has also been displayed.

ACKNOWLEDGEMENT: This work was partically supported by
Fundac$\tilde{a}$o de Amparo $\grave{a}$ Pesquisa do Estado de
S$\tilde{a}$o Paulo (FAPESP) and Conselho Nacional de Desenvolvimento
Cient$\acute{i}$fico e Tecnol$\acute{o}$gico (CNPQ).  B. Wang would also
like to acknowledge the support given by Shanghai Science and Technology
Commission as well as NNSF, China under contract No. 10005004.

\end{document}